\documentclass[printer]{aa}

\usepackage{natbib}
\usepackage{amsmath}
\usepackage{graphicx}
\usepackage{txfonts}
\usepackage{color}
\usepackage{subfloat}

\newcommand{\rg}{\frac{GM}{c^2}}
\newcommand{\tunit}{\frac{GM}{c^3}}
\newcommand{\mdot}{\dot{M}}
\newcommand{\msun}{\mathrm{M}_{\odot}}
\newcommand{\spin}{a_*}
\newcommand{\mbh}{M_{\rm BH}}
\newcommand{\ledd}{L_{\rm Edd}}
\newcommand{\mdotu}{\rm M_\odot yr^{-1}}

\newcommand{\trat}{T_{\rm p}/T_{\rm e}}

\begin{document}

\titlerunning{Scale-invariant radio jets and varying black hole spin}
\title{Scale-invariant radio jets and varying black hole spin}
\author{M. Mo{\'s}cibrodzka\inst{1}\thanks{m.moscibrodzka@astro.ru.nl} \and
  H. Falcke\inst{1} \and S. Noble\inst{2}}
\institute{
Department of Astrophysics/IMAPP,Radboud University,
  P.O. Box 9010, 6500 GL Nijmegen, The Netherlands
\and 
Department of Physics and Engineering Physics,
The University of Tulsa, Tulsa, OK, USA 74104
}
\date{Received June, 2016; accepted x, 2016}
\abstract
    {Compact radio cores associated with relativistic jets are often observed
      in both active galactic nuclei and X-ray binaries. Their radiative
      properties follow some general scaling laws which primarily
        depend on their masses and accretion rates.
      However, it has been suggested that the black
      hole spin can also strongly influence the power and radio flux of these.}
    {Here, we attempt to estimate the dependency of the radio luminosity of
      steady jets launched by accretion disks on black hole mass, accretion
      rate and spin using numerical simulations.}
    {We make use of three-dimensional general relativistic magnetohydrodynamical simulations
      of accretion disks around low-luminosity black holes in which the jet radio
      emission is produced by the jet sheath.}
    {We find that the radio flux increases roughly by a factor of 6 as the
      back hole spin increases from $a_*\approx0$ to $a_*=0.98$. This is comparable
      to the increase in accretion power with spin, meaning that the ratio
      between radio jet and accretion power is hardly changing. Although our
      jet spine power scales as expected for the Blandford-Znajek process, the dependency of
      jet radio luminosity on the black hole spin is somewhat weaker.
      Also weakly rotating black holes can produce visible radio jets. 
      The overall scaling of the radio emission with
      black hole mass and accretion rate is consistent with the scale-invariant
      analytical models used to explain the fundamental plane of black hole
      activity. Spin does not introduce a significant scatter in this model.}
    { The jet-sheath model can describe well resolved accreting systems,
      such as Sgr A* and M87, as well as the general scaling behavior of
      low-luminosity black holes. Hence the model should be applicable to a
      wide range of radio jets in sub-Eddington black holes. The
      black hole spin
      has an effect on the production of visible radio jet, but
        it may not be the main driver to produce visible radio
      jets. An extension of our findings to powerful quasars remains
      speculative.}
\maketitle

\section{Introduction}

Low luminosity active galactic nuclei (LLAGN, defined here as $L_{\rm bol} <
10^{-2} \ledd$ where $L_{\rm bol}$ and $\ledd$ are the bolometric and
Eddington luminosity, respectively) and low-ionization nuclear
emission-line region galaxies (LINERs) often display compact radio
 emissions in
their nuclei (\citealt{falcke:2000}; \citealt{nagar:2000};
\citealt{nagar:2005}).  Their radio cores usually exhibit a flat spectrum (i.e.,
$F_\nu \sim \nu^{-\alpha}$ where $\alpha > -0.5$ for $\nu =$ 1--100~GHz), and are
thought to be associated with the accretion/ejection processes around central
supermassive black holes. High resolution Very Long Baseline Interferometry
(VLBI) observations often reveal that the compact radio cores have morphology
of parsec-scale jets streams (see, e.g., Section 5.4 in \citealt{nagar:2005};
\citealt{markoff:2008}; \citealt{doi:2013} for a sample of LLAGN sources with
compact jets). Compact radio cores are also associated with quasars jets and
blazars, which are much more powerful than LLAGN. These jets are often
associated with high-energy processes such as particle acceleration and X-ray
or $\gamma$-ray emissions (\citealt{grandi:2012}; \citealt{massaro:2015}).

One of the big questions is how do black holes form a radio jet in
the first place? The general picture is that to form a jet one needs
large scale ordered rotating magnetic fields (B-fields). 
However, it is unclear whether the observed jets are powered 
by rotation of an accretion disk in which the B-fields are anchored
or by the spin of the central black hole and rotation of the
B-fields due to frame-dragging effects. It is possible that both
mechanisms operate together.

Spinning black holes have a significant amount of rotational energy. This
energy can be as large as $\sim30$\% of the gravitational energy of
 a supermassive black hole 
($E_{\mathrm{grav}}= \mbh c^2 \approx
10^{62}\,[\mbh/10^8\, \msun]$\,ergs, \citealt{MTW:1977}).  Such energy is
enough to power the radio lobes of a typical radio galaxy. 
The energy removal from the black hole by magnetic fields
was first discussed by \citet{BR:1974}, \citet{Lovelace:1976} and
\citet{BZ:1977}.  Black hole jets associated with the so-called ``Blandford-Znajek
process'' are referred to as Blandford-Znajek jets (hereafter BZ jets). Early
analytical models of BZ jets predicted that the jet power would be
proportional to the square of a black hole rotation, i.e.,
$\mathrm{P_{j}}\sim\Omega_{\mathrm H}^2$ where $\Omega_{\mathrm H}$ is the
angular velocity of the BH event horizon. Later, a number of numerical general
relativistic magnetohydrodynamics (GRMHD) simulations confirmed this
dependency (e.g., \citealt{mckinney:2004}). Some simulations show even
steeper relationship: $\mathrm{P_j} \sim \Omega_{\mathrm H}^{4-6}$ (e.g.,
\citealt{sasha:2010}).  The scaling relation found above, which is often
called the ``spin paradigm'', has been proposed to explain
the diversity of radio jet powers in radio galaxies (FRI-II dichotomy) 
and the radio loud/quiet dichotomy of quasars
\citep{sasha:2010}. The latter is recently being discussed in the context of 
``magnetic flux paradigm" (see e.g., \citealt{sikora:2013}). 

The jet power-spin correlation is difficult to test observationally
because both the power and the spin of the black hole cannot be measured 
{\it directly}. For example, \citet{narayan:2012} estimated the spin of
black holes using a thermal continuum-fitting method in five black
hole binary systems, and found that jet radio luminosity (which by assumption was
associated with the jet power) scales as the square of black hole
spin.  However \citet{russell:2013} casts doubts on these results by
finding no correlation in a larger sample of objects of the same
class. The "spin paradigm" in the context of supermassive black holes in
radio loud galaxies has been addressed by \citet{velzen:2013}, who
found a tight correlation between the optical luminosity of the jet core and
the radio luminosity of jet lobes. The small scatter in their correlation
suggests that a spin of the black hole may not be critical for the
production of jets if one assumes that the black hole spin
distribution function of black holes is uniform. 
\citet{falcke:1995} suggested that jets and disks are symbiotic
features in AGN, and it was shown that the accretion disk luminosity and
jet radio luminosity tightly trace each other over many orders of magnitude in
accretion luminosity and black hole mass (\citealt{rawlings:1991}; 
\citealt{falckemalkan:1995}; \citealt{falcke:1996}).  Later,
\citet{merloni:2003} and \citet{falcke:2004} found a fundamental
plane of black hole activity, connecting X-ray and radio flux of black
holes as a function of mass and accretion rate. Taken at face value,
this suggests that a spin is a subdominant factor in the visible radio jet 
formation and that accretion luminosity and jet radio luminosity 
scale with spin in a comparable manner.

Hence, the role of black hole spin in the production of visible radio
jets is still a highly relevant but unsolved issue. The usual problem is that
the jet radio luminosity is often associated with the jet power. What is then 
the radio jet that we observe in black hole systems and how its luminosity
scales with the black hole spin? Here we attempt to address these
  questions 
from a theoretical point of view, using advanced numerical simulations
of plasma dynamics around spinning black holes. These simulations have
recently become sophisticated enough to produce basic observational
properties of jets (limb-brightening, flat spectrum, size-wavelength
relation, see \citealt{moscibrodzka:2013}; \citealt{chan:2015};
\citealt{moscibrodzka:2016}).  The models to be presented here and the
results obtained from them, however, only concentrate on
low-luminosity supermassive black holes; which have relatively low
accretion rates ($L<10^{-6}\,\ledd$) and optically-thin radiatively
inefficient accretion flows. This significantly simplifies the
necessary physics, as we can largely ignore radiative cooling and
radiation pressure in the dynamical simulations
\citep[e.g.,][]{dibi:2012}. However, it has not been tested yet
whether these models can also reproduce the black hole fundamental plane
scaling and how they respond to changes in the spin parameter.
Fortunately, these GRMHD models can be feasibly combined with a
general relativistic radiative transfer model in a post-processing
manner. This allows one now to examine the electromagnetic emission
from the model from nearly ab initio calculations.

Our jets are produced in time-dependent, fully three-dimensional 
general relativistic magnetohydrodynamic (3D GRMHD) simulations of black hole
accretion flows. The black hole is fed by a magnetized torus seeded
with a single closed loop magnetic field at the equator of the
black hole \citep{moscibrodzka:2014}. 
In such simulations, the BZ-jet naturally emerges, and it
can be seen as a bipolar funnel both above and below the black
hole.  
The funnel regions have ordered large-scale magnetic fields that
allow for energy extraction from BHs if the black hole is rotating fast. 
However, the funnel regions are largely evacuated and radiatively dim. 
In accordance with observations
of high-resolution VLBI images of the M87 jet \citep{hada:2013}, we
identify the main radiating part of the jet as the 
{\em sheath structure} of entrained plasma, which forms around the evacuated
funnels.  In the simulations, the jet sheath is mass loaded in the innermost parts of the
accretion disk, and so its matter content is well known as it is
self-consistently calculated by GRMHD equations. 
Hence, in the jet model proposed here there is no need to, e.g., inject extra
electron/positron pairs to produce radiation from the jet
(\citealt{moscibrodzka:2011}; \citealt{levinson:2011}; \citealt{broderick:2015}).

Moreover, our models naturally fit the detailed observations of the closest
and best studied flat spectrum radio cores, namely Sgr A*
(\citealt{moscibrodzka:2013}; \citealt{moscibrodzka:2014};
\citealt{fraga:2016}) and M87* (\citealt{moscibrodzka:2016}). Our
semi-empirical approach allowed us to constrain unknown parameters, such as
the electron temperatures, in the numerical models by using observations.

Building upon our previous work which uses a fixed black hole spin, we now
present a new suit of numerical models for jets, which are driven from
accretion disks around black holes with a range of spins. For the
first time, we can investigate how the jet radio emission scales with
accretion rate and black hole spin in such models. Also, for the first time we
are able to discuss how the black hole spin affects the scale-invariant
jet models that have been introduced to explain the
fundamental plane of the black hole activity (e.g., \citealt{heinz:2003}).

The paper is organized as follows. In Sect.~\ref{sec:model}, we describe our
simulation setup for the jet formation, and describe how the corresponding
radiation from the jets is computed. In Sect.~\ref{sec:results}, we show
general properties of accretion flow models and we present our studies on how the
radio emission from the jet depends on the black hole spin. We also discuss
the energies and masses carried by different components of the simulations 
(disk, jet sheath, and jet spine). We discuss the result of simulations in the
context of a scale-invariant jet model in Sect.~\ref{sec:dis}. 
  The results are summarized in Sect.\ref{sec:con}.

\section{Method}\label{sec:model}

\subsection{GRMHD simulations}

The dynamics of magnetized plasma near a Kerr black hole is followed by
3D GRMHD simulations. All simulations presented here are carried out using the HARM-3D
code (\citealt{gammie:2003}; \citealt{noble:2009}). The code solves ideal-MHD
equations in  a fixed Kerr metric.

At t=0, the plasma is confined to a geometrically-thick donut-shaped
torus. The plasma density distribution, internal energy, and velocity is
computed using an analytical torus model presented in \citet{fishbone:1976}.
The inner edge of the torus is $r_{in}= 12 GM/c^2$, and the position of the
plasma pressure maximum is $r_{max}=24 GM/c^2$. We follow standard procedures and
seed the initial disk with a sub-thermal magnetic field
($\beta=P_{gas}/P_{mag}=10-100$), the geometry of which aligns with
the iso-density surfaces of the torus (the so-called single loop scenario).

The free parameter of the dynamical simulations is the black hole spin. The spin varies
from $\spin=0.01$ to 0.98, where $\spin=cJ/GM^2$ is the dimensionless angular
momentum of the black hole. Table~\ref{tab:1} lists all considered models with
parameters adopted to set up the computational grid size and resoltuion. We
describe the properties of the GRMHD models in Sect.~\ref{sec:structures}.

\begin{table*}[!t]
\begin{center}
\caption{List of GRMHD models of an accreting black hole with a jet
and the parameters that describe the numerical grid. Column 1: model ID; column 2:
black hole spin; column 3: horizon angular velocity; 
column 4: radius of the innermost stable circular orbit (ISCO); column 5:
Keplerian angular velocity at ISCO; 
columns 6-7: inner edge and location of the pressure maximum of
the disk at t=0 of the GRMHD simulation; 
columns 8-9: inner and outer radius of the simulation domain;
column 10: grid resolution in 3D; column 11: final time of run; column 12:
frequency of data dumps.}~\label{tab:1}
\begin{tabular}{cccccccccccc}
\hline
ID & $a_*$ &$\Omega_{\mathrm H}$& $r_{\mathrm ISCO}$  & $\Omega_{\mathrm
  ISCO}$ &$R_{\mathrm in,t}$ & $R_{\mathrm max,t}$&$R_{\mathrm
  in}$&$R_{\mathrm out}$ & $N1,N2,N3$ &$t_f$
& $\Delta t$\\
&  &$\left[\tunit\right]$& $\left[\rg\right]$  & $\left[\tunit\right]$ &$\left[\rg\right]$ & $\left[\rg\right]$&$\left[\rg\right]$&$\left[\rg\right]$ &  &$\left[\tunit\right]$
& $ \left[\tunit\right]$\\
(1) & (2) & (3) & (4) & (5) & (6) & (7) & (8) & (9) & (10) & (11) & (12)\\
\hline
  A & 0.01  & 0.005& 5.96 & 0.068 & 12&25&  1.1525 &240 & $96,96,64$ & 7000 & 10 \\
  B & 0.2   & 0.101& 5.329& 0.079& 12 &25&  1.1525 &240 & $96,96,64$ & 7000 & 10 \\
  C & 0.5   & 0.267& 4.23 &0.108&12 &24&  1.1525 &240 & $96,96,64$ & 10000 & 10 \\
  D & 0.7  & 0.408& 3.15&0.157& 12 &24&  1.1525 &240 & $96,96,64$ & 8000 & 10 \\
  E & 0.9375& 0.695& 2.04&0.26&12 &24&  1.1525 &240 & $96,96,64$ & 10000 & 10 \\
  F & 0.96875&0.776&1.75&0.3&12 &24&  1.1525 &240 & $96,96,64$ & 8100 & 10 \\
  G & 0.98438&0.837& 1.55&0.34&12 &24&  1.1525 &240 & $96,96,64$ & 10000 & 10 \\
\hline
\end{tabular}
\end{center}
\end{table*}

\subsection{Radiative transfer models}

To calculate the electromagnetic emission from the GRMHD simulations,
we use a ray-tracing radiative transfer method. In this method, the
radiative transfer equation is integrated along null geodesics
from the vicinity of the black hole to an observer. 
The radiative transfer includes synchrotron emission and
self-absorption from a relativistic, thermal population of electrons
described by the Maxwell-J{\"u}ttner distribution function. We note that
most AGN show evidence for a non-thermal power-law distribution with a
low-energy cutoff. A thermal distribution plus a non-thermal tail
would be a more appropriate description in this case. However, this
would introduce an additional free parameter, and for the self-absorbed
flat-spectrum part of jets, it is anyway the low-energy peak that
dominates the emission. In fact, in simple analytic Blandford-K\"onigl
type models of flat-spectrum radio cores, thermal and non-thermal
distribution functions produce the same scaling behavior (e.g., 
\citealt{falcke:2002}).

With this distribution function, synchrotron emission maps at a given
frequency are calculated. 
The integration of intensity over the map gives us
the model total flux. By producing maps at various frequencies, we
construct spectral energy distribution (SED) of synchrotron emission
from the model. Radiative transfer calculations are carried out with
the general relativistic radiative transfer code {\tt mibothros}
described by \citet{noble:2007}. Our numerical code has been tested
extensively by the authors. We have recently compared our current
code with the synchrotron maps produced by an independent code
{\tt grtrans} \citep{dexter:2016}, and confirmed that the emitted
radiation is independent of the radiative transfer scheme.

We have to assume an electron temperature as they are not
self-consistently computed in our simulations. The usual problem here
is that typical GRMHD simulations only provide the proton temperature,
$T_{\rm p}$. The electron temperature is therefore usually
parameterized by a coupling ratio between proton and
electron temperatures ($\trat$). In \citet{moscibrodzka:2013}, we suggested that
the coupling ratio is most likely different in the low-magnetized disk
compared to the tenuous, highly-magnetized jet sheath. Indeed, fitting
of the spectral energy distributions (SED) of Sgr A* and M87 suggests
that the electron-proton coupling is much lower in the disk, compared
to the jet. The disk flow is therefore described as a two-temperature
plasma, as commonly assumed for advection dominated/radiatively
inefficient accretion flow (ADAF/RIAF) solutions
\citep{narayan:1998}, while the jet is described as a hot, single
temperature electrons. The latter is justified by the almost constant
brightness temperatures seen in compact jets and may be due to
continuous re-acceleration/re-heating of electrons along the jet or
due to thermal conductivity along the magnetic field lines, balancing
some of the adiabatic losses.  Notice that the electron temperature
prescriptions used here and in our previous work are now naturally
produced in the axisymmetric extended-GRMHD models that follow the
electron and proton temperatures from nearly first-principles
(\citealt{ressler:2015}; \citealt{foucart:2016}).

In our models, the boundary between the disk and jet regions is
defined by using the hydrodynamical Bernoulli parameter, $-h u_t=1.02$
where $h$ is gas enthalpy, and $u_t$ is the covariant time component of the
gas four-velocity. 
The fiducial value of the Bernoulli parameter typically corresponds to a
jet plasma bulk velocities $v > 0.2c$, as measured by an observer at
infinity. This is, e.g., consistent with subluminal speeds detected at
the base of, e.g., M87 jet \citep{hada:2016}. 
Such a definition consistently associates the magnetically dominant
outflow region with the jet component and makes the jet sheath radio bright. 
We neglect radiation from the highly
magnetized and evacuated jet spine because the jet-spine temperatures
and matter content are arbitrarily set by the numerical floor
quantities (fluid density and internal energy floors) that are
necessary for numerical stability of the GRMHD code. 

Here, we are not fitting models to a specific set of observed SEDs; hence, we adopt a
constant value, $\trat=20$ for the disk, which would fit typical SEDs
well. The default jet electron temperature
is set as $\Theta_{e,jet}=kT_e/m_ec^2=20$ in electron rest mass energy
units. The latter is also motivated by fitting models to Sgr A* spectrum 
(e.g., \citealt{moscibrodzka:2014}, \citealt{chan:2015}, \citealt{gold:2016}). 
$\trat$ and $\Theta_{e,jet}$ are fixed parameters in all the radiative
transfer calculations.

Finally, to completely define the radiative transfer problem we have to scale
the GRMHD simulation to a specific source. This requires providing the central black
hole mass (sets the typical length scale) and mass of the
accretion disk (sets the accretion rate onto the black hole). 
Our fiducial model assumes a mass of the central black hole in a LLAGN to be
$M_{\mathrm BH}=10^8 \msun$. We consider a system in which the black hole is accreting
at a rate $\dot{m} =\dot{M}/\dot{M}_{\mathrm Edd} \approx 10^{-5}$, where 
the Eddington accretion rate is defined as usual, i.e.,
$\dot{M}_{\mathrm Edd}=L_{\mathrm Edd}/0.1c^2= 4 \pi G M_{\mathrm BH} m_p
c/(\sigma_{\mathrm TH}
0.1c^2) \approx 2.2 (M_{\mathrm BH} / 10^8 M_{\mathrm \odot}) \,\, {\rm [\mdotu]}$. 
In Sect.~\ref{sec:dis}, where we discuss the scale-invariant 
jet models, the parameters $M_{\mathrm BH}$ and
$\dot{m}$ are varied. 

\section{Results}\label{sec:results}

\begin{figure}
\begin{center}
\includegraphics[width=0.4\textwidth,angle=0]{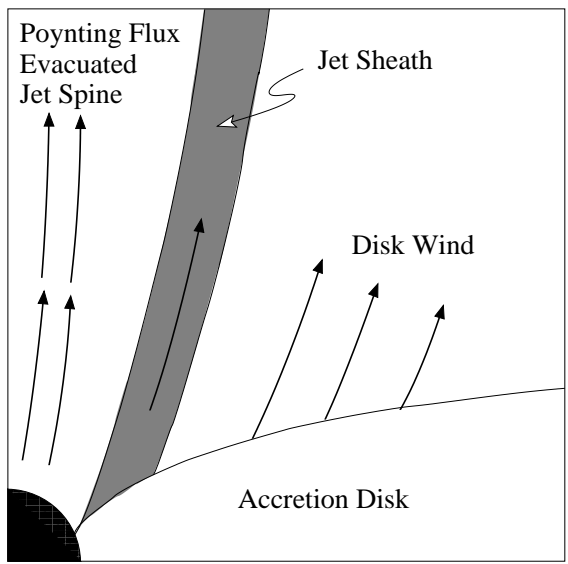}
\caption{ A schematic diagram of the geometry and the components
  of our 3-D GRMHD simulations of accretion flow with a jet. 
  Only one hemisphere of the model is
  shown. The black hole is located at the lower left corner of the plot.
  The figure is adopted after \citet{hawley:2006}.}
\label{fig:sketch}
\end{center}
\end{figure}

Fig.~\ref{fig:sketch} presents a schematic diagram of our 3-D GRMHD
simulations. During the simulations, three regions with various
physical conditions are formed near the black hole. In the disk
region, located at the equatorial plane of the grid, the plasma is
weakly magnetized and turbulent, and it is accreting onto the black
hole.  The jet-spine is the highly magnetized region above the black
hole poles where the plasma content is negligible, and the force-free
conditions prevail. The evacuated jet-spine is surrounded by the jet
sheath. We will find here that this general picture does not change
with the black hole spin.

\subsection{Structures of jets and their accretion flows}\label{sec:structures}

Before presenting  the radiative properties of the jets as a function of black hole
spin, we first examine some of the properties of the disk, the jet spine, and the jet sheath in
models with various black hole spins (models A through F in Table~\ref{tab:1}).

\begin{figure}
\begin{center}
\includegraphics[width=0.48\textwidth,angle=0]{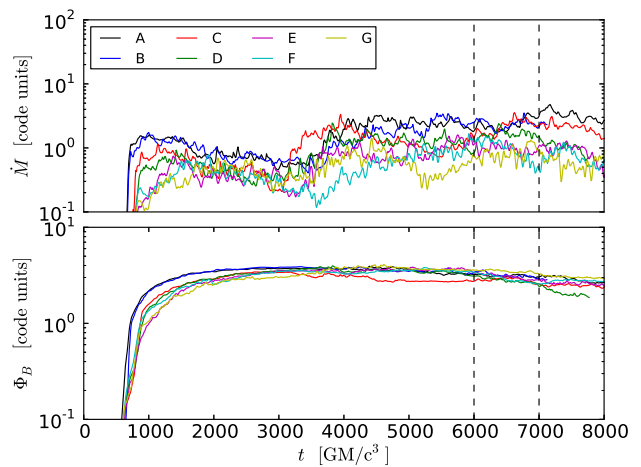}
\caption{Top panel: mass accretion rate onto the black hole
as a function of time in models with varying black hole spin. Bottom panel: the 
poloidal magnetic field flux accumulated near the black hole event horizon as
a function of time for various black hole spins. All quantities are shown in
dimensionless (code) units. }
\label{fig:mdotBdot}
\end{center}
\end{figure}

Fig.~\ref{fig:mdotBdot} (top panel) shows the evolution of the mass accretion
rate, where the colors distinguish the models 
with various black hole spins. The accretion
rate is calculated using a standard definition, i.e., 
\begin{equation}
\dot{M}_{net}=\int_\theta \int_\phi \rho_0 u^r(r=r_H) dA_{\theta\phi},
\end{equation}
where $\rho_0$ is the rest mass density, $u^r$ is the radial component of the
plasma four-velocity, $r_H=1 + \sqrt{1-\spin^2}$ is the BH horizon radius, and
$dA_{\theta\phi} = \sqrt{-g} d\theta d\phi$ is the element of the surface area
in the Kerr metric with $\sqrt{-g}$ being the metric determinant.  Evidently, the
mass accretion rate (presented in Fig.~\ref{fig:mdotBdot} in arbitrary units)
in all models saturates and fluctuates around unity. The accretion rate is
variable due to the inhomogeneous structure of the turbulent accretion
disk. The accretion rate averaged over later times of the simulation is
 approximately the same for all black hole spins.

In the BZ model, the power of a black hole jet depends not only
  on the black hole spin but also on
the magnetic field flux accumulated near the black hole horizon, $P_j \propto \spin^2
\Phi_B^2$. The absolute magnetic field flux that is being accumulated near the
black hole horizon as the simulation evolves is computed using
\begin{equation}
\Phi_{B,r_H}=  \frac{1}{2} \int_{\pi/2}^{\pi} \int_\phi  |B^r| dA_{\theta\phi}, 
\end{equation}
where $B^r$ is the radial component of the magnetic field vector.
Fig.~\ref{fig:mdotBdot} (lower panel) shows that in all models similar amount
of dimensionless magnetic flux has been accumulated at the black hole horizon 
  regardless of the black hole spin.  
 Here our purpose was to show that a free parameter of the
model such as $\Phi_B$ is actually fixed, and so for a constant accretion 
rate onto the black hole, any eventual differences in
the jet power or jet radio luminosity would be due to altering the black hole
spin. Notice that if we change the accretion rate onto the black hole (by scaling the
simulation density), we automatically increase the magnetic field flux
at the black hole because the strength of the magnetic field increases.
If the mass of the black hole is increased then the flux near the black hole decreases
because we decrease the B field strength.

In our simulations, the jet spine is the evacuated region above the black hole
poles and it is equivalent to the BZ jet. For a non-rotating black hole, the
jet-spine also forms but no energy extraction from the black hole occurs.
The electromagnetic power of the jet-spine is defined as $P_{\rm j} = \int_{\rm jet} \sqrt{-g} dx^2 dx^3
F_{\rm E}^{(EM)}$, where $F_E^{(EM)}$ is the dimensionless electromagnetic radial
energy flux defined as $F_{\rm E}^{(EM)}=(T^r_t)^{(EM)} = b^2 u^r u_t - b^r
b_t $. The jet powers, scaled to a system with the fiducial black
hole mass $M_{\rm BH}=10^8 \msun$ and accretion rate $\dot{m}=10^{-5}$, are listed
in Table~\ref{tab:2}.

We find that the electromagnetic energy is extracted from the black
hole for $a_* \geqslant 0.5$, and the jet power increases with the
black hole spin according to $P_j \propto \Omega_{\mathrm H}^{2}$ 
(where $\Omega_{\mathrm H}(\spin)=\spin/2r_H$ and
 where $r_H$ is the radius of the black hole event horizon). These findings
are in agreement with the BZ model and with what has been found in similar simulations by
other authors. In our models, the efficiency of the
 black hole rotational energy extraction, defined as $\eta_j=P_{\rm j}/\dot{M}c^2$ is at most 4\%,
where $P_j$ and $\dot{M}$ are the time-averaged electromagnetic
luminosity inside of the evacuated funnel and accretion rate through
the black hole horizon, respectively. Typically, the models studied
here are called SANE in the current jargon (which stand for Standard
And Normal Evolution), having low-power jets.  For comparison, jets
produced in so-called Magnetically Arrested Disks (MADs) are more
efficient because in MADs $\Phi_{B,r_H}$ is typically about 10 times
larger compared to our values (e.g., \citealt{sadowski:2013}). Consequently,
the energy extraction efficiency in MADs can reach 140\% \citep{sasha:2011}.

\begin{table}
\begin{center}
\caption{The electromagnetic power of the jet-spine $P_{\mathrm j}$ as a
  function of black hole spin $a_*$. The jet power in c.g.s. units 
    is given by multiplying the dimensionless jet power calculated in the GRMHD 
    simulation by the jet power unit given by $P_{\mathrm unit} = 9 \times10^{46}
    \dot{m} (M_{\mathrm BH}/10^8 {\mathrm M_{\odot}}) {\mathrm [ergs \, s^{-1}]}$.
    For $M_{\mathrm BH}=10^8 M_{\odot}$ and $\dot{m}=10^{-5}$ the $P_{\mathrm unit} = 
    9 \times10^{41} {\mathrm [ergs \, s^{-1}]}$.}\label{tab:2}
\begin{tabular}{cccc}
\hline
%model ID &$\spin$ & $\Omega_{\mathrm H}$ & $P_{\rm j} / 9\times10^{41} {\mathrm [ergs \, s^{-1}]}$ \\
model ID &$\spin$ & $\Omega_{\mathrm H}$ & $P_{\rm j} {\mathrm [P_{\mathrm unit}]}$ \\
\hline
A &$0.01$&0.005 & $0.00192$ \\
B &$0.2$ &0.101& $3.34\times 10^{-4}$\\
C &$0.5$ &0.267& $-0.00184$\\
D &$0.7$ &0.408& $-0.00995$\\
E &$0.94$&0.695 & $-0.02$\\
F &$0.96$&0.776 & $-0.043$\\
G &$0.98$&0.837 & $-0.034$\\
\hline
\end{tabular}
\end{center}
\end{table}

\begin{figure*}
\begin{center}
\includegraphics[width=1.0\textwidth,angle=0]{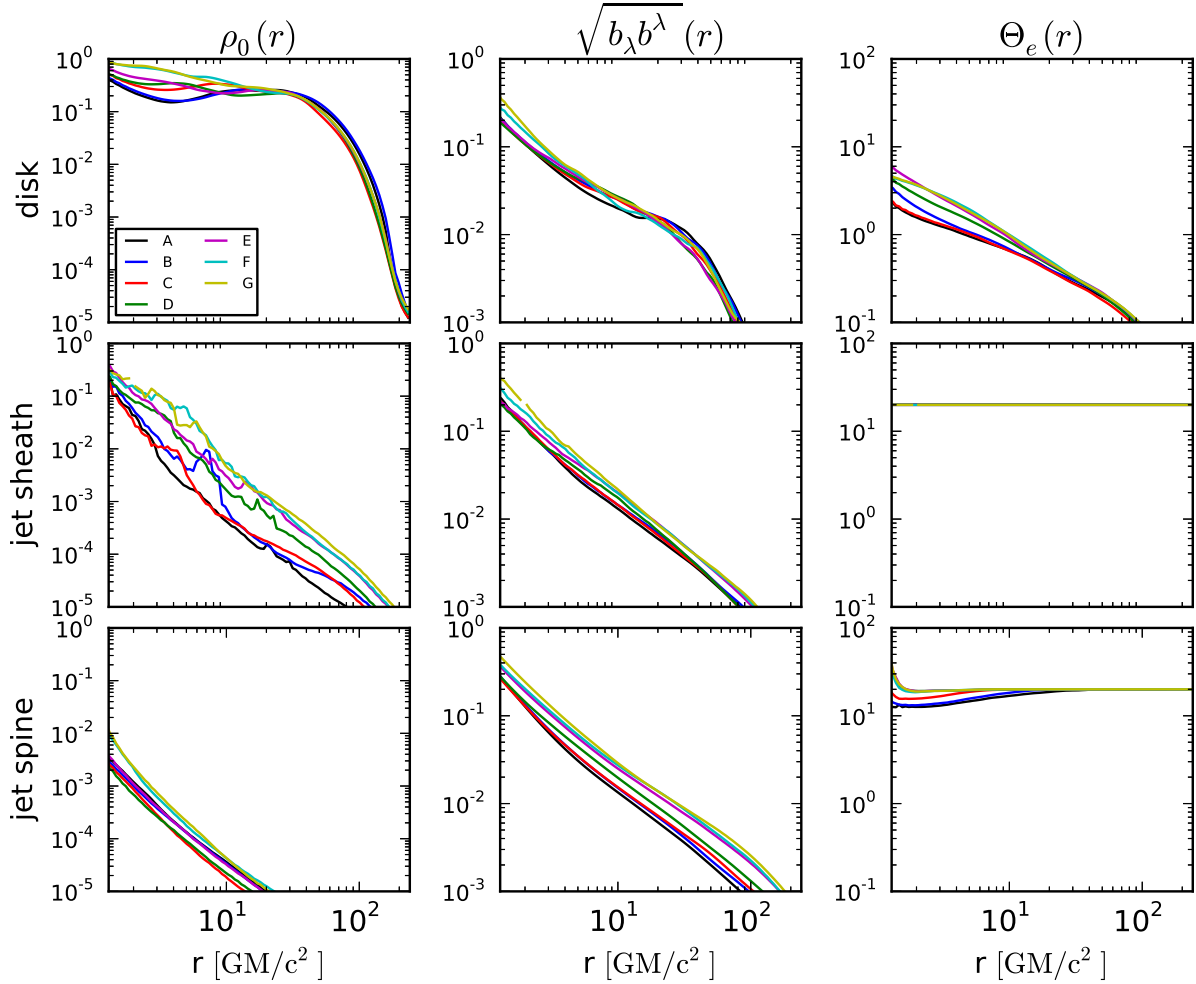}
\caption{Shell- and time-averaged profiles of rest mass density $\rho_0$ (left column), magnetic
  field strength $\sqrt{b_\lambda b^\lambda}$ (middle column), and electron
  temperature $\Theta_e$ (right column) in
  the disk (top row), jet sheath (middle row), and jet spine (bottom row) for
  models with various black hole spins. All quantities are in dimensionless
  code units. See text (Sect.~\ref{sec:structures}) for discussion.}
\label{fig:prof1}
\end{center}
\end{figure*}

Next, we present time- and shell-averaged radial profiles 
of various scalar  quantities in  the simulations 
that are important for the radiative transfer
calculations  since they are used in the synchrotron
emissivity and absorptivity functions. 
The averaging is defined as follows:
\begin{equation}
\left< q(r)\right> = \frac{1}{\Delta t} \int_{t_{min}}^{t_{max}} \frac{
  \int_{0}^{\pi}  \int_{0}^{2\pi}  
q(r, \theta, t) \sqrt{-g} d\theta d\phi}
{ \int_{0}^{\pi} \int_{0}^{2\pi}    \sqrt{-g} d\theta d\phi} dt 
\end{equation}
where $q$ is a scalar quantity of interest. We time-average the quantities
over later times of the simulation when the solutions relax from initial
conditions and reach a quasi-stationary state.
 The integration limits $t_{\mathrm min}$=6000 ${\mathrm GM/c^3}$ and
$t_{\mathrm max}=$7000 ${\mathrm GM/c^3}$ are
indicated in Fig.~\ref{fig:mdotBdot} as two vertical dashed
lines. In Fig.~\ref{fig:prof1}, we show the time-averaged radial profiles of plasma density, magnetic
field strength and assumed electron temperature (where
electron temperature is expressed as $\Theta_e=k_bT_e/m_ec^2$). These
quantities are shown for the accretion disk region, in the jet sheath and in the jet spine regions. 

We first discuss the radial structures of the jet sheath.  Examining
Fig.~\ref{fig:prof1}, it is evident that in the jet sheath the plasma
density decreases with radius as $\rho_0\sim r^{-2}$ and the magnetic
fields strength decreases with distance as $B\sim r^{-1}$. This is the
same dependence as the one derived in the simple conical outflows used in
analytical modeling of jets in the past. As predicted by
\citet{blandford:1979} and \citet{falcke:1995}, such jets are expected
to produce a nearly flat spectral index in their SEDs when the electron
distribution function along the jet is fixed. This is in agreement
with our findings here and with \citet{moscibrodzka:2013} and
\citet{falcke:2000}. Interestingly, we also find that the density of
the jet sheath depends on the black hole spin. This is in agreement
with findings by \citet{hawley:2006} (see their Table~2.) who show
that the mass outflow through the jet sheath increases with black hole
spin. The spin of the black hole does not affect the radial dependency
of the density in the jet-sheath but only the normalization constant. This has interesting
implications in the scale-invariant jet models, as discussed in
Sect.~\ref{sec:dis}.

Fig.~\ref{fig:prof1} also indicates the same radial dependencies of $\rho_0$
and B with radius in the jet spine. However, the mass loading of the jet spine is negligible
compared to the jet sheath; hence, most of the visible radiation will be
generated by the jet sheath. Finally, in the disk the radial dependencies of plasma
density and magnetic fields are flatter compared to that in the jet. The
electron temperatures in the disk become sub-relativistic beyond $10 GM/c^2$ and
so the accretion disk neither emits nor absorbs synchrotron radiation.

\subsection{Radio luminosity of a jet}\label{sec:radio}

Through out this paper, we assume the inclination angle of a jet as
$i=60^\circ$ when calculating its radio emission. Although not shown here, the
radio flux computed for a higher inclination angle decreased by a factor of
two at most, and vice versa for a lower inclination model. This trend is
roughly in agreement with the change in the flux due to the relativistic
Doppler factor $D=\Gamma^{-1}(1-v/c \cos(i))^{-1}$ where $\Gamma=1/\sqrt{1-v^2/c^2}$ 
and $i$ is the inclination angle. Hence, for a mildly relativistic jet
(v=0.2 -- 0.5c), the flux is expected to change by no more than a factor of two.

\begin{figure*}
\begin{center}
\includegraphics[width=0.48\textwidth,angle=0]{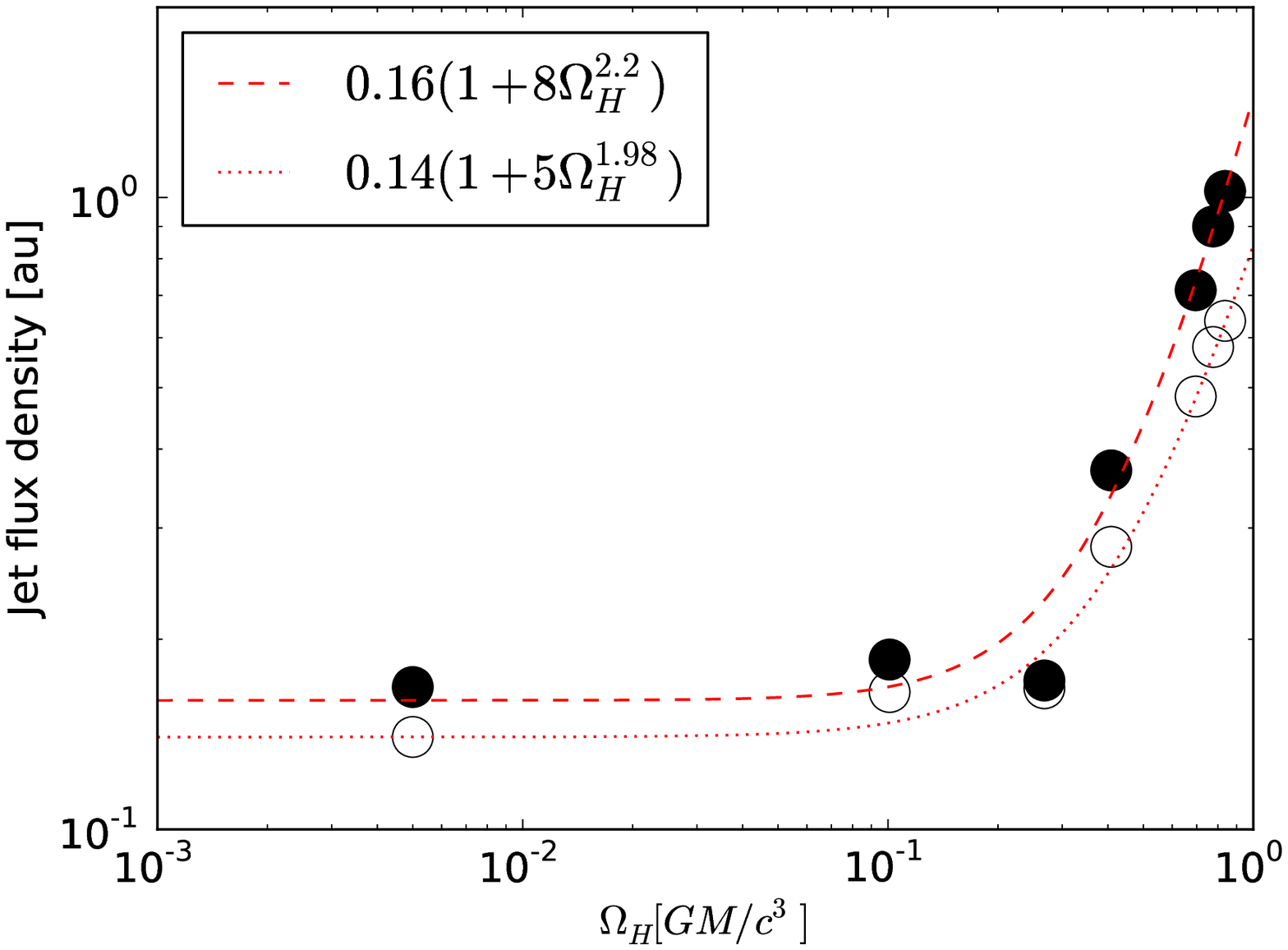}
\includegraphics[width=0.48\textwidth,angle=0]{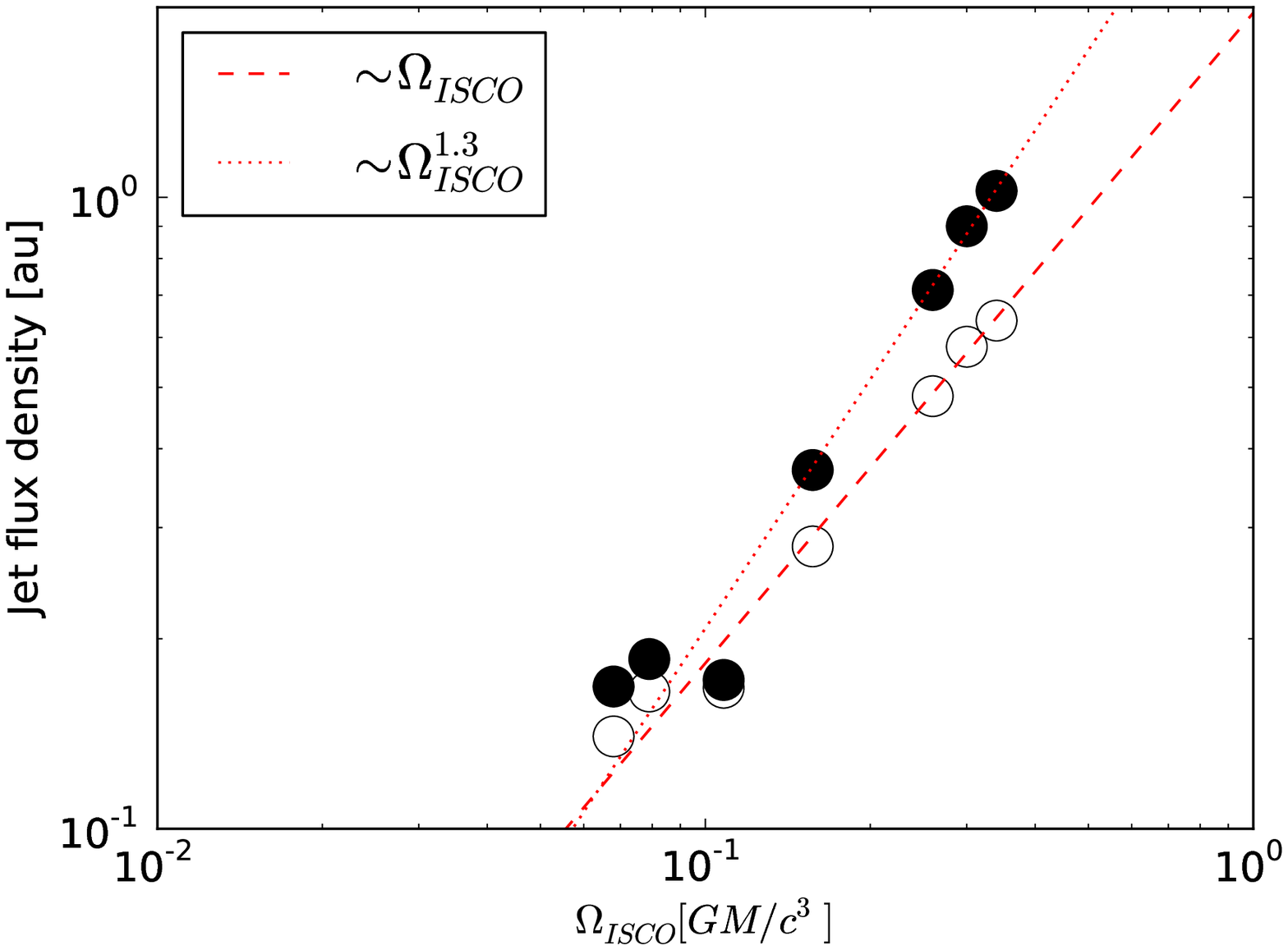}
\caption{  The radio flux densities of models plotted 
  as a function of the black hole event
  horizon angular velocity (left panel) and as a function of the angular
  velocity of the matter in the disk at the ISCO (right panel). 
  We show the radio flux densities at two frequencies; $\nu_1=$86
  (filled circles) $\nu_2=$43 (open circles) GHz at which the SEDs are close to flat and optically
  thick.  The flux densities are given in arbitrary units since
  they depend on the distance to
  the source that is not specified here. 
  The emission is computed for a fixed mass of the black hole and a fixed
  accretion rate. The lines show the best fits to the synthetic data points.
 The best fit parameters along with the form of the fitting
  function are given in Table~\ref{tab:3}.}
\label{fig:spin}
\end{center}
\end{figure*}

Fig.~\ref{fig:spin} displays the radio flux emitted by the jet
measured at two frequencies by a distant observer as a function of
black hole spin.  The radio fluxes shown are time-averages of jet
radio emission produced within $\Delta t = 1000 GM/c^3$ of the
simulation, when the accretion flow and jet are fully relaxed from the
initial conditions. For the assumed $\dot{m}\approx10^{-5}$ and
$\mbh=10^8 \msun$, the jet synchrotron flux at $\nu_1$=86 and
$\nu_2=$43 GHz are produced in the optically thick part of the jet
base.  Two frequencies are shown to demonstrate that the model
spectral index $\alpha$, defined as $F_\nu \sim \nu^{-\alpha}$,
matches the typical observational value of compact, self-absorbed
jets. In our models, for all black hole spins, we find 
$\alpha \approx -0.5$, which is often called an `inverted spectrum`.

Fig.~\ref{fig:spin} (left panel) shows the radio 
  flux densities plotted against the black hole
event horizon angular velocity $\Omega_{\mathrm H}$.
Solid and dashed lines in
this figure are the parametric fit of the flux density as a
  function of $\Omega_{\mathrm H}$. The dependency of the
jet radio flux density on $\Omega_{\mathrm H}$ can be described approximately with
$f(\Omega_{\mathrm H})=a(1 + b\Omega_{\mathrm H}^{p})$. We present the best least square fit
values of the free parameters $a$, $b$, and $p$ in Table~\ref{tab:3}.
These dependencies are less steep than the BZ
jet power dependency on spin: $P_{j}\sim \Omega_{\mathrm H}^{2-6}$.  In other
words even {\it for a nearly zero spin the jet radio emission persists}.

\begin{table}
\begin{center}
\caption{The parameters for the best fit models for the radio flux
  densities (measured at the frequencies: $\nu_1=86$ and $\nu_2=43$ GHz). The
  fitting functions are assumed to have the following forms:
  $f(\Omega_H)=a(1+b \Omega_H^p)$ where $\Omega_H$ is the angular velocity of the
  black hole horizon ($\Omega_H$) or $f(\Omega_{ISCO})=a\Omega_{ISCO}^p$ 
  where $\Omega_{ISCO}$ is the angular velocity at
  the innermost stable circular orbit. The fitting parameters are $a$, $b$, and $p$.
}\label{tab:3}
\begin{tabular}{ccccc}
\hline
$f(\Omega)$ & $\nu$ &$a$ & $b$ & $p$\\
\hline
$f(\Omega_H)$ &$\nu_1$ & $0.16\pm0.02$ & $7.99\pm1.25$ & $2.28\pm0.29$\\
$f(\Omega_H)$ &$\nu_2$ & $0.14\pm0.013$& $4.99\pm0.51$ & $1.98\pm0.22$\\
$f(\Omega_{ISCO})$& $\nu_1$ & $1.9\pm0.14$  & $\cdots$  &$1.03\pm0.05$\\
$f(\Omega_{ISCO})$& $\nu_2$ & $4.24\pm0.45$  & $\cdots$  & $1.31\pm0.08$\\
\hline
\end{tabular}
\end{center}
\end{table}

In the right panel in Fig.~\ref{fig:spin} the same jet radio flux
densities are plotted against the angular frequency of the inner most stable circular
(ISCO) orbit $\Omega_{\mathrm ISCO}=(r_{\mathrm ISCO}^{3/2}+\spin)^{-1}$
 the formula
for the angular velocity of a Keplerian disk near a rotating black hole). We have
 found that our disks indeed have the angular
  velocities similar to those of the Keplerian angular
velocities. For convenience, the values of $\Omega_{\mathrm ISCO}$ are displayed in
Table~\ref{tab:1}. The dependency of the radio flux on $\Omega_{\mathrm ISCO}$ can be
described by a trivial function of $a\,\Omega_{\mathrm ISCO}^p$.  The best fit
parameters $a$ and $p$ are summarized in Table~\ref{tab:3}. The slightly different values
of $p$ parameter for different observing frequencies are due to limited size of
our jet which is limited by the size of out computational grid and due to 
the curvature in the spectrum. We argue that the two
  chosen frequencies bridge the
valid range of the model. The observed radio flux is approximately
proportional to the Keplerian angular frequency of matter at the ISCO.
We argue that the increase in the jet radio flux 
is a result of a higher mass content of
the jet sheath as other relevant variables remain roughly unchanged.

\subsection{Mass and energy fluxes carried by the jet}

It is interesting to investigate the mass and energy fluxes in the
accretion disk and in the two-component jet because the dissipation of
energy and radiation in AGN might be also produced by, e.g., the jet
interacting with the environment of the black hole. It is important to
find out whether the mass and energy fluxes depend on the black hole
spin and whether the black hole spin could play a role in the black hole
feedback.

The radial profiles of mass and total energy flows are defined as:
\begin{equation}
\dot{M}(r) =  \int_\theta \int_\phi (\rho_0 u^r) dA_{\theta\phi}
\end{equation}
and
\begin{equation}
\dot{E}_{tot}(r) =  \int_\theta \int_\phi (-T^r_t) dA_{\theta\phi},
\end{equation}
where $T^r_t=(\rho_0+\gamma u + b^2) u^r u_t - b^r b_t$ is a component of the
stress-energy MHD tensor that represents the energy of the plasma and the
magnetic field transported in the radial direction, and 
$dA_{\theta\phi}=\sqrt{-g} d\theta d\phi$ is an area element in the $\theta-\phi$
plane. The symbols in the tensor expression have the following meanings: $\rho_0$ is
the plasma rest-mass density, $\gamma$ is the adiabatic index, $u$ is the
specific internal energy of plasma, $b^\mu$ is the magnetic field four-vector
defined in the frame comoving with the plasma, and $u^\mu$ is the plasma four-velocity.
The negative (positive) value of $\dot{M}$ and $\dot{E}_{tot}$ indicates inflow (outflow)
of the mass/energy from the system. We also calculate the energy flux reduced
by the rest mass energy, i.e., 
\begin{equation}
\dot{E}(r) = \int_\theta \int_\phi (-T^r_t - \rho_0 u^r) dA_{\theta\phi}.
\end{equation}
Here, $\dot{E}$ is the sum of kinetic, magnetic and thermal energies.  

\begin{figure*}
\begin{center}
\includegraphics[width=1.0\textwidth,angle=0]{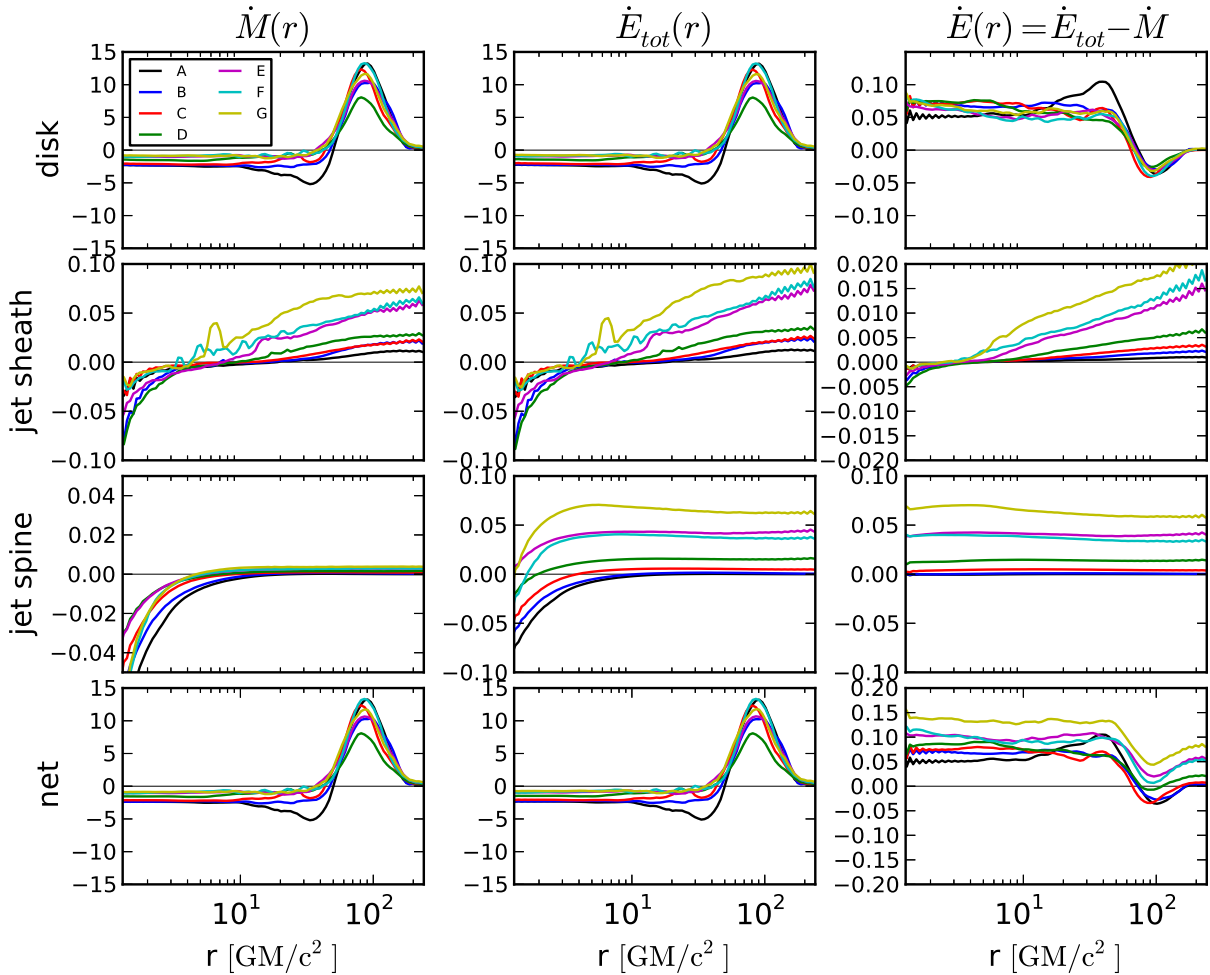}
\caption{Shell- and time-averaged radial fluxes of mass, total energy, and
  energy minus rest mass energy ( $\dot{M}$, $\dot{E}_{tot}$, and
    $\dot{E}$ from left to right, respectively) in the disk, jet sheath, jet
  spine, and the whole regions for models with various
  black hole spins.}
\label{fig:4a}
\end{center}
\end{figure*}

Fig.~\ref{fig:4a} displays $\dot{M}$, $\dot{E}_{tot}$, and $\dot{E}$
in the disk, jet sheath, and jet spine regions in models A to F. 
All energy fluxes are given in dimensionless units used in the numerical code. 
In the disk region (first row in Fig.~\ref{fig:4a}), the mass is inflowing towards
the black hole only starting at $r\sim50 GM/c^2$ which corresponds roughly to
the pressure maximum of the torus. The massive outflow beyond this radius is
caused by the torus wind. Interestingly, the energy in the disk region is
positive. This indicates that the energy is removed from the inner parts of the disk.
The properties of radial mass and energy fluxes seem independent of the black
hole spin as one would naively expect.

The radial mass and energy fluxes in the jet sheath are pointing
outwards and they
increase with the black hole spin. The mass flux in the jet is dominated by
the jet sheath which can expel up to 5\% of the mass accreting onto the
black hole. In the jet spine, the mass flux is defined by 
  the numerical floor values.

The net mass flux (i.e., integrated over the all angles) 
is controlled by the accretion disk, however the net
energy radial flux is pointing outwards at all radii and the
dependency on the black hole spin is evident. Interestingly, both the
jet spine and jet sheath contribute to the total energy flux outflow,
indicating that both jet components may play some role in the black hole
feedback which is then somewhat spin-dependent.

Finally, we are interested in which component of the energy is dominating the energy
in disk, jet sheath, and jet spine regions.  Following definitions in \citet{mckinney:2012}
(their Eq. 6)  the radial energy flux $\dot{E}(r)$ can be further 
split into magnetic ($\dot{E}_{\mathrm mag}$), kinetic ($\dot{E}_{\mathrm kin}$), and thermal
($\dot{E}_{\mathrm th}$) energies. It is straight forward to
find that the radial components of these energy fluxes are
\begin{equation}
\dot{E}_{\mathrm mag}=B^2 u^r u_t - b^r b_t
\end{equation}
\begin{equation}
\dot{E}_{\mathrm kin}=\rho_0 u^r (u_t +1) 
\end{equation}
\begin{equation}
\dot{E}_{\mathrm th} = \gamma u u^r u_t.
\end{equation}

\begin{figure*}
\begin{center}
\includegraphics[width=1.0\textwidth,angle=0]{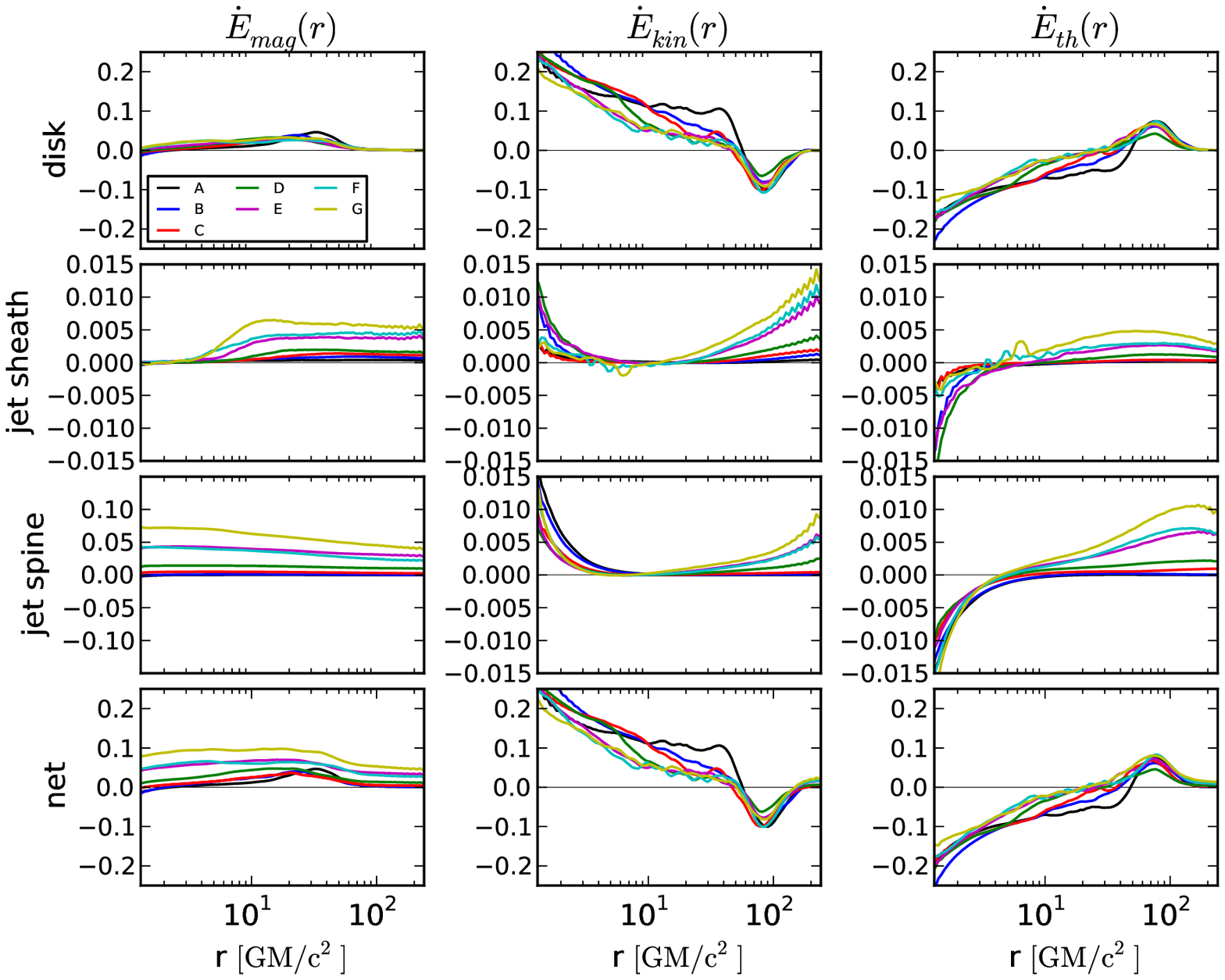}
\caption{Shell- and time-averaged radial fluxes of mass, total energy, and
  energy without rest mass energy included ($\dot{E}_{\mathrm mag}$, $\dot{E}_{\mathrm kin}$,
$\dot{E}_{\mathrm th}$ from left to right, respectively) in the disk, jet sheath, jet
  spine, and the whole regions in models with various black hole spins.}
\label{fig:5a}
\end{center}
\end{figure*}

Figs.~\ref{fig:5a} shows that the magnetic, kinetic and thermal (associated
mainly with heavy protons) energies of the three regions in our
models. The inner ($r < 50 GM/c^2$) accretion disk is dominated by the
kinetic and thermal energy with weak contribution from the magnetic
energy. This is expected in a weakly magnetized plasma. It is evident
that in the jet sheath  all three
energies increase as the black hole spin increases.  The energies are
roughly in equipartition around $r \sim 50 GM/c^2$ and the jet sheath
acceleration rate increases with increasing spin. The net magnetic
energy of the jet is predominantly within the jet spine.
It is worth noting that the radiative losses of the jet sheath 
are comparable to the rotational energy that is being liberated from the
black hole in the BZ process that operates in the jet spine.
For example, for black hole spin $a_*=0.94$ and fixed
$\dot{m}\approx10^{-5}$ and $\mbh=10^8 \msun$,
the total luminosity of the system (luminosity integrated over frequencies and averaged over
angles) is approximately $6 \times 10^{40} [ergs \, s^{-1}]$ which is close
the BZ power of the jet spine which is $3 \times 10^{40} [ergs \, s^{-1}]$
(see Table~\ref{tab:2}).

\section{Scale-invariant jet models}~\label{sec:dis}

In this section we discuss our results in the context of
scale-invariant jet models (\citealt{falcke:1995},
\citealt{heinz:2003}).  The scale-invariant models were introduced to
unify jets physics in black hole systems with various masses and
accretion rates, i.e., AGN or X-ray binaries, and to explain the
relationship between the black hole mass, radio luminosity and X-ray
luminosity of the object --- so-called 'fundamental plane of black holes'
(\citealt{merloni:2003}; \citealt{falcke:2004}).  Since our models are
fully relativistic, for the {\it first time} we can test these analytic
models, and inspect the effect of black hole spin for these
models. On the other hand, one could hope to constrain more parameters in
the GRMHD simulations by comparing them to the fundamental plane of
black hole activity and the empirical scaling laws.

The scale-invariant jet model provides a relationship between the observed jet
radio flux $F_{\nu}$ (where $\nu$ is the observing frequency) and the global
parameters of the system: the mass of a central black hole $\mbh$, black hole
spin ($\spin$), and the mass accretion rate $\mdot$ in Eddington units.
In the scale-invariant jet model, any dynamically relevant variable $f$ (e.g., plasma
density or magnetic field strength) is a product of two decoupled functions:
\begin{equation}
f(\mbh,\mdot,\spin,r) = \phi_f(\mbh,\mdot,\spin) \, \psi_f(r'c^2/GM_{\rm BH})
\end{equation}
where $\phi_f$ describes the dependency of the variable on the central engine
mass, accretion rate, and spin and $\psi_f$ describes the spatial dependency
of variable $f$ on the similarity variable $r'c^2/GM_{\rm BH}$, which is simply the
distance from the center in units of gravitational radii (that scales with
mass of the black hole).

The synchrotron emission from the jet depends on the electron number
density $n_e$ and the magnetic field strength, $B$, along the jet.
Jets launched by the ADAF models, which are what we effectively have in our
numerical simulations, are mechanically cooled jets (by adiabatic
decompression), as evident from Fig.~\ref{fig:prof1}. Here only
$\phi_{n_e}$ depends on the black hole spin, while other relevant
quantities $\psi_{n_e}$, $\phi_B$, and $\psi_B$ depend only weakly on
it. Consequently, our GRMHD simulations produce the following density and
magnetic field dependencies:
\begin{equation}
\phi_{n_e} \propto f(\spin) \left(\frac{\dot{M}}{M_{\rm BH}^2}\right)
\end{equation}
and 
\begin{equation}
\phi_B \propto \left(\frac{\dot{M}}{M_{\rm BH}^2}\right)^{1/2}.
\end{equation}
Based on our numerical models, we find $f(\spin) \sim \Omega_{\mathrm ISCO}$
(Sect.~\ref{sec:radio}). The radial dependencies in the jet sheath 
  (Sect.~\ref{sec:structures}), as evident in Fig.~\ref{fig:prof1} are:
\begin{equation}
\psi_{n_e} \propto \left(\frac{r'c^2}{GM_{\rm BH}}\right)^{-2}
\end{equation}
\begin{equation}
\psi_{B} \propto \left(\frac{r'c^2}{GM_{\rm BH}}\right)^{-1}
\end{equation}
We find that the above dependencies (except the spin dependency) roughly agree
with the conical jet model in \citet{blandford:1979} or \citet{falcke:1995}, and 
with the scale-invariant jet model from ADAFs (\citealt{heinz:2003}; \citealt{yuan:2002}).

We can therefore use the latter model to obtain the general scalings
for jet optically thick emission. We plug in our $\phi_{n_e}$, $\phi_B$,
$\psi_{n_e}$ and $\psi_B$ derived from GRMHD simulations to Eqs. 9, 10, and
12 in \citet{heinz:2003}. This results in similar relationships which are now
modified by black hole spin, that enters the radiative transfer integral by scaling the jet
particle number density. We find that the jet radio flux ($F_{\nu}$) scales with mass and
accretion rate as:
\begin{equation}
\frac{\partial ln (F_\nu)}{\partial ln (M_{\rm BH})} = \frac{17}{12} -  \frac{\alpha}{3}
\end{equation}
and
\begin{equation}
\frac{\partial ln (F_\nu)}{\partial ln (\dot{m})}=\frac{17}{12} + \frac{2 \alpha}{3}
\end{equation}
where $\alpha$ is the spectral index.
Interestingly, because of the logarithmic derivatives, any
dependency on the spin cancels out, thus our formulae are identical to those
provided by \citet{heinz:2003}.  The black hole spin does not affect the
scalings of flux with black hole mass and accretion rate but it affects the
observed flux (their Eq. 8) which is also evident in our radiative
transfer models (Fig.~\ref{fig:spin}).

Next, we compare the theoretical scalings to the results of fully
relativistic radiative transfer calculations. In our radiative
transfer simulations, we find that for small changes of $\dot{m}$ and
$M_{\rm BH}$,  $F_\nu \propto \dot{m}^{1.1}$ and $F_\nu \propto M_{\rm BH}^{1.5}$ (or equivalently
$F_\nu \propto \dot{M}^{1.1}$ and $F_\nu \propto M_{\rm BH}^{0.4}$) and
$F_\nu$ scalings is independent of the black
hole spin. Notice that our numerical experiments show that the
observed jet radio spectrum in general is not flat
(the spectral index $\alpha\neq0$, but rather $-0.5$). 
Hence, the numerical dependency on the accretion rate (expressed in dimensionless Eddington
units) turn out to be in excellent agreement with \citet{heinz:2003} prediction
$17/12+2\alpha/3=1.08$ assuming a spectral slope $\alpha_\nu=-0.5$,
and the dependency on the mass of the black hole $17/12-\alpha/3=1.58$,
is also very good. Our value of $\alpha\neq0$ is most probably due to the
non-conical shape of jet near the black hole and the limited extent
of the simulation domain.

Finally, the dependency of the flux on the black hole spin parameter is
described by the function $f(\spin)$ which enters the radiative transfer integral as a jet
density scaling constant. The optically thick part of the radio emission from
the jet will be simply proportional to $f(\spin)$.

\section{Summary}\label{sec:con}

We have shown that 3-D GRMHD simulations of low-luminosity supermassive black holes
can reproduce the compact flat-spectrum radio emission for various black
hole spins ranging from $\spin \approx 0.0$ to $\spin=0.98$. 
In our simulations the visible radio jet is launched from an optically-thin 
radiatively inefficient accretion flow, and is identified as the plasma
outflowing in a thin layer surrounding a pure Poynting flux
outflow. In such a scenario, the observed radio flux generated by the jet 
increases with black hole spin. We argue that the 
increase in the radio flux is caused by more efficient mass loading of the jet sheath.
The radio flux detected by the distant observer changes by a factor
of less than 10, from a low to a high spin values.

This indeed seems to be a large factor; however, we note that in
standard thin disks, the accretion power scales as $r_{\mathrm ISCO}^{-1}$, which can 
vary by a factor of 6 between spinning and non-spinning black holes. This is
simply because the matter falls deeper into  
the potential well and acquires more energy for a smaller ISCO.
{\it Hence, the jet radio luminosity normalized to accretion power does not change as a function of spin.}

We find that the radio flux scales with black hole mass and accretion
rate remarkably similar to what is found in analytic scale-invariant
jet models,  in which the spin is a minor factor. This validates some of the
basic assumptions of the jet-disk symbiosis and fundamental plane
pictures, that require a relatively low scatter between accretion disk
luminosities and radio fluxes or jet powers.

The present model of jet radio emission is not only motivated by
observational characteristics of LLAGN such as flat-to-inverted radio
spectra on the self-absorbed side which are typical for astrophysical
jets but also motivated by a couple of other theoretical
findings. For example, the
jet sheath emission can fit broadband observational data of Sgr~A*
(e.g., \citealt{moscibrodzka:2014}; \citealt{chan:2015}; \citealt{gold:2016}).  
We have also recently discussed in detail the application of the bright jet
sheath models to model the observations of jet core in M87 galaxy (see
\citealt{moscibrodzka:2016}), in which the sheath is actually resolved in
VLBI observations. In addition, our models of jet emission can be
tested in detail using the millimeter and sub-millimeter VLBI observations that will directly image
black holes and the surrounding plasma in Sgr~A* and M87 in the near future.
 The model presented in this work may be applicable to a wide
  range of black hole jet-disk systems with different masses and accretion
  rates because we find essentially the same scaling laws in our jet-sheath
  models as in the analytical scale-invariant jet model.

Finally, our model still has a number of uncertainties that may
affect the presented results. Neglecting radiative cooling may or may
not affect the scaling laws mentioned above as one goes to higher
accretion rates. The dependency of radio fluxes may be affected by,
e.g., relatively low resolution of the numerical grid; our arbitrary
definition of the jet and disk zones or initial conditions that are not 
completely free of spin effects. The most important
simplification however is the very robust prescription for emitting
particles - the Maxwell-J{\"u}ttner distribution function 
  parametrized with a constant electron temperature. This
  simplification is also the reason why we have not discussed here the high energy emission
from the optically thin parts of the jet. The electrons in jets
certainly form a power-law distribution, which has its largest effect
at frequencies higher than radio. The high energy emission could be
entirely dominated by the power-law tail of electrons or
inverse-Compton emission produced by the power-law electrons in the
jet. If so, the latter could possibly also explain, e.g., the observed
X-ray/radio correlation in black hole binaries and AGNs
(\citealt{merloni:2003}; \citealt{falcke:2004};
\citealt{plotkin:2012}). This should be explored in future
studies.

\begin{acknowledgements}
We acknowledge support from the ERC Synergy Grant “BlackHoleCam – Imaging the Event
Horizon of Black Holes” (Grant 610058). Authors thank
BlackHoleCam collaborators and R. Kurosawa for comments on the manuscript.
\end{acknowledgements}

%\bibliographystyle{aa}
%\bibliography{local}

\begin{thebibliography}{52}
\expandafter\ifx\csname natexlab\endcsname\relax\def\natexlab#1{#1}\fi

\bibitem[{{Blandford} \& {K{\"o}nigl}(1979)}]{blandford:1979}
{Blandford}, R.~D. \& {K{\"o}nigl}, A. 1979, \apj, 232, 34

\bibitem[{{Blandford} \& {Rees}(1974)}]{BR:1974}
{Blandford}, R.~D. \& {Rees}, M.~J. 1974, \mnras, 169, 395

\bibitem[{{Blandford} \& {Znajek}(1977)}]{BZ:1977}
{Blandford}, R.~D. \& {Znajek}, R.~L. 1977, \mnras, 179, 433

\bibitem[{{Broderick} \& {Tchekhovskoy}(2015)}]{broderick:2015}
{Broderick}, A.~E. \& {Tchekhovskoy}, A. 2015, \apj, 809, 97

\bibitem[{{Chan} {et~al.}(2015){Chan}, {Psaltis}, {{\"O}zel}, {Narayan}, \&
  {Sa{\c d}owski}}]{chan:2015}
{Chan}, C.-K., {Psaltis}, D., {{\"O}zel}, F., {Narayan}, R., \& {Sa{\c
  d}owski}, A. 2015, \apj, 799, 1

\bibitem[{{Dexter}(2016)}]{dexter:2016}
{Dexter}, J. 2016, ArXiv e-prints

\bibitem[{{Dibi} {et~al.}(2012){Dibi}, {Drappeau}, {Fragile}, {Markoff}, \&
  {Dexter}}]{dibi:2012}
{Dibi}, S., {Drappeau}, S., {Fragile}, P.~C., {Markoff}, S., \& {Dexter}, J.
  2012, \mnras, 426, 1928

\bibitem[{{Doi} {et~al.}(2013){Doi}, {Kohno}, {Nakanishi}, {Kameno}, {Inoue},
  {Hada}, \& {Sorai}}]{doi:2013}
{Doi}, A., {Kohno}, K., {Nakanishi}, K., {et~al.} 2013, \apj, 765, 63

\bibitem[{{Falcke} \& {Biermann}(1995)}]{falcke:1995}
{Falcke}, H. \& {Biermann}, P.~L. 1995, \aap, 293

\bibitem[{{Falcke} \& {Biermann}(1996)}]{falcke:1996}
{Falcke}, H. \& {Biermann}, P.~L. 1996, \aap, 308, 321

\bibitem[{{Falcke} \& {Hehl}(2002)}]{falcke:2002}
{Falcke}, H. \& {Hehl}, F. 2002, {The Galactic Black Hole}

\bibitem[{{Falcke} {et~al.}(2004){Falcke}, {K{\"o}rding}, \&
  {Markoff}}]{falcke:2004}
{Falcke}, H., {K{\"o}rding}, E., \& {Markoff}, S. 2004, \aap, 414, 895

\bibitem[{{Falcke} {et~al.}(1995){Falcke}, {Malkan}, \&
  {Biermann}}]{falckemalkan:1995}
{Falcke}, H., {Malkan}, M.~A., \& {Biermann}, P.~L. 1995, \aap, 298, 375

\bibitem[{{Falcke} \& {Markoff}(2000)}]{falcke:2000}
{Falcke}, H. \& {Markoff}, S. 2000, \aap, 362, 113

\bibitem[{{Fishbone} \& {Moncrief}(1976)}]{fishbone:1976}
{Fishbone}, L.~G. \& {Moncrief}, V. 1976, \apj, 207, 962

\bibitem[{{Foucart} {et~al.}(2016){Foucart}, {Chandra}, {Gammie}, \&
  {Quataert}}]{foucart:2016}
{Foucart}, F., {Chandra}, M., {Gammie}, C.~F., \& {Quataert}, E. 2016, \mnras,
  456, 1332

\bibitem[{{Fraga-Encinas} {et~al.}(2016){Fraga-Encinas}, {Mo{\'s}cibrodzka},
  {Brinkerink}, \& {Falcke}}]{fraga:2016}
{Fraga-Encinas}, R., {Mo{\'s}cibrodzka}, M., {Brinkerink}, C., \& {Falcke}, H.
  2016, \aap, 588, A57

\bibitem[{{Gammie} {et~al.}(2003){Gammie}, {McKinney}, \&
  {T{\'o}th}}]{gammie:2003}
{Gammie}, C.~F., {McKinney}, J.~C., \& {T{\'o}th}, G. 2003, \apj, 589, 444

\bibitem[{{Gold} {et~al.}(2016){Gold}, {McKinney}, {Johnson}, \&
  {Doeleman}}]{gold:2016}
{Gold}, R., {McKinney}, J.~C., {Johnson}, M.~D., \& {Doeleman}, S.~S. 2016,
  ArXiv e-prints

\bibitem[{{Grandi} {et~al.}(2012){Grandi}, {Torresi}, \& {on behalf of the
  FERMI-LAT collaboration}}]{grandi:2012}
{Grandi}, P., {Torresi}, E., \& {on behalf of the FERMI-LAT collaboration}.
  2012, ArXiv e-prints

\bibitem[{{Hada} {et~al.}(2016){Hada}, {Kino}, {Doi}, {Nagai}, {Honma},
  {Akiyama}, {Tazaki}, {Lico}, {Giroletti}, {Giovannini}, {Orienti}, \&
  {Hagiwara}}]{hada:2016}
{Hada}, K., {Kino}, M., {Doi}, A., {et~al.} 2016, \apj, 817, 131

\bibitem[{{Hada} {et~al.}(2013){Hada}, {Kino}, {Doi}, {Nagai}, {Honma},
  {Hagiwara}, {Giroletti}, {Giovannini}, \& {Kawaguchi}}]{hada:2013}
{Hada}, K., {Kino}, M., {Doi}, A., {et~al.} 2013, \apj, 775, 70

\bibitem[{{Hawley} \& {Krolik}(2006)}]{hawley:2006}
{Hawley}, J.~F. \& {Krolik}, J.~H. 2006, \apj, 641, 103

\bibitem[{{Heinz} \& {Sunyaev}(2003)}]{heinz:2003}
{Heinz}, S. \& {Sunyaev}, R.~A. 2003, \mnras, 343, L59

\bibitem[{{Levinson} \& {Rieger}(2011)}]{levinson:2011}
{Levinson}, A. \& {Rieger}, F. 2011, \apj, 730, 123

\bibitem[{{Lovelace}(1976)}]{Lovelace:1976}
{Lovelace}, R.~V.~E. 1976, \nat, 262, 649

\bibitem[{{Markoff} {et~al.}(2008){Markoff}, {Nowak}, {Young}, {Marshall},
  {Canizares}, {Peck}, {Krips}, {Petitpas}, {Sch{\"o}del}, {Bower}, {Chandra},
  {Ray}, {Muno}, {Gallagher}, {Hornstein}, \& {Cheung}}]{markoff:2008}
{Markoff}, S., {Nowak}, M., {Young}, A., {et~al.} 2008, \apj, 681, 905

\bibitem[{{Massaro} {et~al.}(2015){Massaro}, {Harris}, {Liuzzo}, {Orienti},
  {Paladino}, {Paggi}, {Tremblay}, {Wilkes}, {Kuraszkiewicz}, {Baum}, \&
  {O'Dea}}]{massaro:2015}
{Massaro}, F., {Harris}, D.~E., {Liuzzo}, E., {et~al.} 2015, \apjs, 220, 5

\bibitem[{{McKinney} \& {Gammie}(2004)}]{mckinney:2004}
{McKinney}, J.~C. \& {Gammie}, C.~F. 2004, \apj, 611, 977

\bibitem[{{McKinney} {et~al.}(2012){McKinney}, {Tchekhovskoy}, \&
  {Blandford}}]{mckinney:2012}
{McKinney}, J.~C., {Tchekhovskoy}, A., \& {Blandford}, R.~D. 2012, \mnras,
423,
  3083

\bibitem[{{Merloni} {et~al.}(2003){Merloni}, {Heinz}, \& {di
  Matteo}}]{merloni:2003}
{Merloni}, A., {Heinz}, S., \& {di Matteo}, T. 2003, \mnras, 345, 1057

\bibitem[{{Misner} {et~al.}(1977){Misner}, {Thorne}, \& {Wheeler}}]{MTW:1977}
{Misner}, C.~W., {Thorne}, K.~S., \& {Wheeler}, J.~A. 1977, {Gravitation.}

\bibitem[{{Mo{\'s}cibrodzka} \& {Falcke}(2013)}]{moscibrodzka:2013}
{Mo{\'s}cibrodzka}, M. \& {Falcke}, H. 2013, \aap, 559, L3

\bibitem[{{Mo{\'s}cibrodzka} {et~al.}(2016){Mo{\'s}cibrodzka}, {Falcke}, \&
  {Shiokawa}}]{moscibrodzka:2016}
{Mo{\'s}cibrodzka}, M., {Falcke}, H., \& {Shiokawa}, H. 2016, \aap, 586, A38

\bibitem[{{Mo{\'s}cibrodzka} {et~al.}(2014){Mo{\'s}cibrodzka}, {Falcke},
  {Shiokawa}, \& {Gammie}}]{moscibrodzka:2014}
{Mo{\'s}cibrodzka}, M., {Falcke}, H., {Shiokawa}, H., \& {Gammie}, C.~F. 2014,
  \aap, 570, A7

\bibitem[{{Mo{\'s}cibrodzka} {et~al.}(2011){Mo{\'s}cibrodzka}, {Gammie},
  {Dolence}, \& {Shiokawa}}]{moscibrodzka:2011}
{Mo{\'s}cibrodzka}, M., {Gammie}, C.~F., {Dolence}, J.~C., \& {Shiokawa}, H.
  2011, \apj, 735, 9

\bibitem[{{Nagar} {et~al.}(2005){Nagar}, {Falcke}, \& {Wilson}}]{nagar:2005}
{Nagar}, N.~M., {Falcke}, H., \& {Wilson}, A.~S. 2005, \aap, 435, 521

\bibitem[{{Nagar} {et~al.}(2000){Nagar}, {Falcke}, {Wilson}, \&
  {Ho}}]{nagar:2000}
{Nagar}, N.~M., {Falcke}, H., {Wilson}, A.~S., \& {Ho}, L.~C. 2000, \apj, 542,
  186

\bibitem[{{Narayan} {et~al.}(1998){Narayan}, {Mahadevan}, {Grindlay},
    {Popham},
  \& {Gammie}}]{narayan:1998}
{Narayan}, R., {Mahadevan}, R., {Grindlay}, J.~E., {Popham}, R.~G., \&
  {Gammie}, C. 1998, \apj, 492, 554

\bibitem[{{Narayan} \& {McClintock}(2012)}]{narayan:2012}
{Narayan}, R. \& {McClintock}, J.~E. 2012, \mnras, 419, L69

\bibitem[{{Noble} {et~al.}(2009){Noble}, {Krolik}, \& {Hawley}}]{noble:2009}
{Noble}, S.~C., {Krolik}, J.~H., \& {Hawley}, J.~F. 2009, \apj, 692, 411

\bibitem[{{Noble} {et~al.}(2007){Noble}, {Leung}, {Gammie}, \&
  {Book}}]{noble:2007}
{Noble}, S.~C., {Leung}, P.~K., {Gammie}, C.~F., \& {Book}, L.~G. 2007,
  Classical and Quantum Gravity, 24, S259

\bibitem[{{Plotkin} {et~al.}(2012){Plotkin}, {Markoff}, {Kelly},
    {K{\"o}rding},
  \& {Anderson}}]{plotkin:2012}
{Plotkin}, R.~M., {Markoff}, S., {Kelly}, B.~C., {K{\"o}rding}, E., \&
  {Anderson}, S.~F. 2012, \mnras, 419, 267

\bibitem[{{Rawlings} \& {Saunders}(1991)}]{rawlings:1991}
{Rawlings}, S. \& {Saunders}, R. 1991, \nat, 349, 138

\bibitem[{{Ressler} {et~al.}(2015){Ressler}, {Tchekhovskoy}, {Quataert},
  {Chandra}, \& {Gammie}}]{ressler:2015}
{Ressler}, S.~M., {Tchekhovskoy}, A., {Quataert}, E., {Chandra}, M., \&
  {Gammie}, C.~F. 2015, \mnras, 454, 1848

\bibitem[{{Russell} {et~al.}(2013){Russell}, {Gallo}, \&
  {Fender}}]{russell:2013}
{Russell}, D.~M., {Gallo}, E., \& {Fender}, R.~P. 2013, \mnras, 431, 405

\bibitem[{{S{\c a}dowski} {et~al.}(2013){S{\c a}dowski}, {Narayan}, {Penna},
    \&
  {Zhu}}]{sadowski:2013}
{S{\c a}dowski}, A., {Narayan}, R., {Penna}, R., \& {Zhu}, Y. 2013, \mnras,
  436, 3856

\bibitem[{{Sikora} \& {Begelman}(2013)}]{sikora:2013}
{Sikora}, M. \& {Begelman}, M.~C. 2013, \apjl, 764, L24

\bibitem[{{Tchekhovskoy} {et~al.}(2010){Tchekhovskoy}, {Narayan}, \&
  {McKinney}}]{sasha:2010}
{Tchekhovskoy}, A., {Narayan}, R., \& {McKinney}, J.~C. 2010, \apj, 711, 50

\bibitem[{{Tchekhovskoy} {et~al.}(2011){Tchekhovskoy}, {Narayan}, \&
  {McKinney}}]{sasha:2011}
{Tchekhovskoy}, A., {Narayan}, R., \& {McKinney}, J.~C. 2011, \mnras, 418, L79

\bibitem[{{van Velzen} \& {Falcke}(2013)}]{velzen:2013}
{van Velzen}, S. \& {Falcke}, H. 2013, \aap, 557, L7

\bibitem[{{Yuan} {et~al.}(2002){Yuan}, {Markoff}, \& {Falcke}}]{yuan:2002}
{Yuan}, F., {Markoff}, S., \& {Falcke}, H. 2002, \aap, 383, 854

\end{thebibliography}

\end{document}